\documentclass[10pt,a4paper]{article}
\usepackage[cp1251]{inputenc}
\usepackage[russian]{babel}
\inputencoding{cp1251}
\usepackage{amsmath}
\usepackage{amsthm}
\usepackage[usenames]{color}
\usepackage[dvips]{graphicx}
\usepackage{varioref}
\usepackage{amssymb}
\usepackage[plainpages]{hyperref}
\def\bb{\begin{equation}}
\def\ee{\end{equation}}

\def\ve{\varepsilon}

\title{{\bf Система нелинейных осцилляторов с диссипацией. Начальный этап авторезонанса.  \thanks{Работа выполнена при поддержке гранта РФФИ 11-01-91330-ННИО-а} }}

\author{С. Глебов \thanks{Уфимский государственный Нефтяной технический университет({\tt glebskie@gmail.com})} \\
Н. Тарханов \thanks{Институт математики, Потсдамский университет ({\tt tarkhanov@math.uni-potsdam.de})}         }

\date{\null}

\begin{document}

\maketitle

\begin{abstract}
В работе исследуется влияние диссипации на порог авторезонанса в возмущенной нелинейной  системе связанных осцилляторов. Приведены явные асимптотические формулы и численное моделирование эффектов возникающих на начальном этапе обрушения авторезонасного роста решения.
\end{abstract}

\section{Введение}

Системы связанных нелинейных осцилляторов часто используются для описания колебательных процессов в различных областях знаний. Наиболее интересным эффектом в таких системах является резонанс -- возрастание амплитуды колебаний. Резонансное взаимодействие может возникать как между собственными модами, так и с внешней силой, если такая имеется. Медленное изменение частот может приводить к новым эффектам, к захвату в резонанс, при котором частота внешней силы приходит к резонансному значению и в дальнейшем остается около него. Эффекты такого типа принято называть авторезонансом \cite{McMillan}-\cite{Veksler2}.

Мы будем рассматривать систему возмущенных слабо нелинейных связанных осцилляторов с диссипацией. Собственные частоты постоянны и относятся друг к другу как 1:2. Это случай так называемого параметрического резонанса. В работе \cite{klg} показано, что за счет малого возмущения в такой системе без диссипации можно получить решения большой амплитуды, но при этом амплитуда возмущения должна превышать некоторое пороговое значение. Результаты о пороге авторезонанса в системах без диссипации были получены как численно \cite{Friedland1}-\cite{Friedland2}  так и аналитически \cite{Kalyakin1}-\cite{Kalyakin2}, \cite{klg}.

Имеются также результаты о влиянии диссипации на порог авторезонанса \cite{Friedland3},\cite{Friedland4}, \cite{GKS}. В работах \cite{SKKS1}, \cite{SKKS2}, \cite{GKT1}, \cite{GKT2}  получены оценки максимального значения амплитуды колебаний момента срыва решения. 

Детальное исследование авторезонансных  систем с диссипацией приведено в работах авторов \cite{GKT1}, \cite{GKT2}.

Целью данной работы является получение аналитического описания влияния диссипации на авторезонансные процессы в системах нелинейных осцилляторов. Оценка величины диссипации, при которой в системе возможен авторезонанс за счет внешнего возмущения.

\section{Постановка задачи и редукция к модельной системе}

Рассмотрим систему слабо связанных осцилляторов с диссипацией
\begin{eqnarray}
x''+\ve\nu_1 x'+ \omega^2 x &=& \ve \alpha_1 xy +\ve (\gamma\exp\{i\varphi+c.c. \}),\nonumber\\ 
y''+\ve\nu_2 y'+ 4\omega^2 y &=& \ve \alpha_2 x^2. \label{sysOfOscillators} 
\end{eqnarray}
здесь $\ve$ малый положительный параметр,  $\varphi=(\omega+\ve\alpha \tau)\theta,\ \tau=\ve\theta$.

Параметры в системе (\ref{sysOfOscillators}) имеют следующий смысл: $\omega$ -- собственная частота первого осциллятора, $\alpha_1, \alpha_2$ -- параметры нелинейной связи осцилляторов, $\nu_1,\nu_2$ -- коэффициенты диссипации, $\gamma$ -- амплитуда внешнего возмущения, $\alpha$ -- производная от расстройки частоты внешнего возмущения по медленному времени $\tau$.

Решение (\ref{sysOfOscillators}) будем искать в комплексной форме
\bb
x={\cal A}(\tau)\exp\{i\omega\theta\}+c.c.,\quad y={\cal B}(\tau)\exp\{2i\omega\theta\}+c.c.. \label{subs1}
\ee
Подставляя (\ref{subs1}) в  (\ref{sysOfOscillators}) и оставляя только слагаемые порядка $\ve$ получаем 
\begin{eqnarray*}
{\cal A}'(\tau) &=& -\frac{i\alpha_1}{2\omega}{\cal A}^*{\cal B} - \frac{i\gamma}{2\omega}\exp\{i\alpha\tau^2\}-\frac{\nu_1}{2} {\cal A},\\
{\cal B}'(\tau) &=& -\frac{i\alpha_2}{4\omega}{\cal A}^2 - \frac{\nu_2}{2} {\cal B}.
\end{eqnarray*}

Подстановка
$$
{\cal A} = a(\tau)\exp\{i\alpha\tau^2\}+c.c.,\quad {\cal B} = b(\tau)\exp\{2i\alpha\tau^2\}+c.c.
$$
приводит к системе
\begin{eqnarray*}
a'+2i\alpha\tau a + \frac{\nu_1}{2}a&=&-\frac{i\alpha_1}{2\omega}a^*b - \frac{i\gamma}{2\omega}, \\
b'+4i\alpha\tau b + \frac{\nu_2}{2}b&=&-\frac{i\alpha_2}{4\omega}a^2.
\end{eqnarray*}

Масштабированием независимой переменной $\tau$ и амплитуд $a, b$ часть параметров можно сделать равными единице. Замена переменных
$$
\begin{array}{ccc}  
\tau=\chi t, & a=\lambda A, & b=\kappa B, \\
& & \\
\kappa=\displaystyle\frac{\omega\sqrt{\alpha}}{\alpha_1}, & \lambda =\displaystyle\sqrt{\frac{\alpha}{\alpha_1\alpha_2}},  & \chi^2=\displaystyle\frac{1}{\alpha}
\end{array}
$$
приводит к системе уравнений главного резонанса
\begin{eqnarray}
A'(t) &=& -i\Big(2tA + \frac{1}{2}A^*B+f\Big)-\mu_1A   , \nonumber\\
B'(t) &=& -i\Big(4tB + \frac{1}{4}A^2 \Big)-\mu_2B, \label{mainSys}
\end{eqnarray}
\noindent где 
$$
\mu_1=\displaystyle\frac{\nu_1}{2\sqrt{\alpha}},\qquad 
\mu_2=\displaystyle\frac{\nu_2}{2\sqrt{\alpha}},\qquad  f=\displaystyle\frac{\gamma\sqrt{\alpha_1\alpha_2}}{2\alpha\omega^2}.
$$

\section{Численное моделирование и мотивация постановки задачи}

\begin{figure}
\includegraphics[width=14cm,height=8cm]{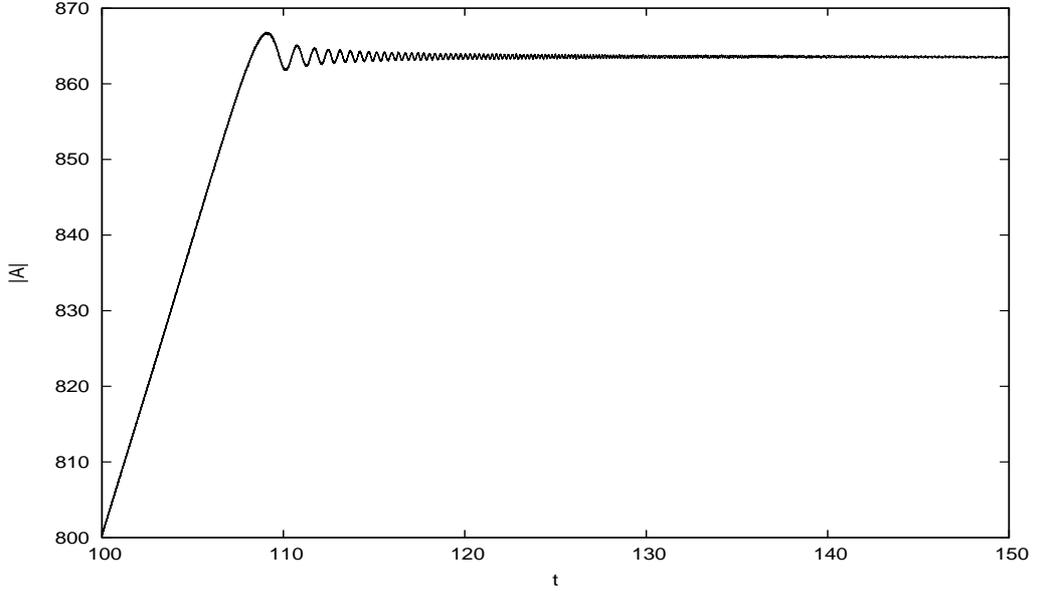}
\caption{График $|A(t)|$ решения (\ref{mainSys}) при $\mu_1=\mu_2=0$ и значении  $f=11.9$ меньше порогового значения $f_*=12$.}
\label{fig1}
\end{figure}

\begin{figure}
\includegraphics[width=14cm,height=8cm]{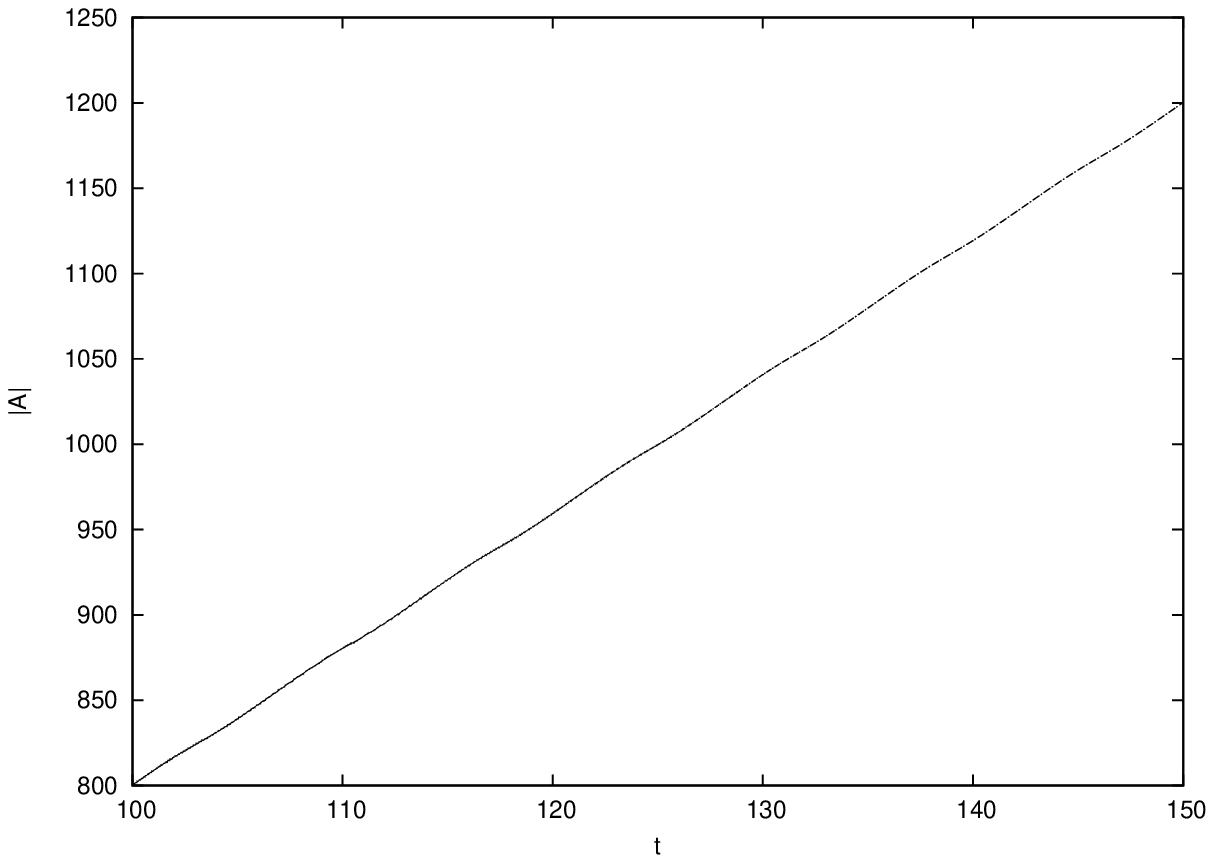}
\caption{График $|A(t)|$  решения (\ref{mainSys}) при $\mu_1=\mu_2=0$ и значении $f=12.1$ больше порогового значения $f_*=12$.}
\label{fig11}
\end{figure}

В работе \cite{klg} показано, что для системы (\ref{mainSys}) при $\mu_1=\mu_2=0$ имеется пороговое значение амплитуды возмущения $f_*=12$. Если амплитуда возмущения $f$ превышает это пороговое значение, то в системе происходит захват в резонанс. Если же амплитуда возмущения меньше, то захвата не происходит. На Рис. \ref{fig1},\ref{fig11} приведены результаты численного моделирования этого эффекта. Численный счет для системы (\ref{mainSys}) при $\mu_1=\mu_2=0$ проводился методом Рунге--Кутты 4-го порядка. Величина шага сетки составляла $10^{-4}$. График на Рис.\ref{fig1}, соответствуют значению $f=11.9 < f_*$, а график на Рис. \ref{fig11} ---  значению $f=12.1 > f_*$. 

\begin{figure}
\includegraphics[width=14cm,height=8cm]{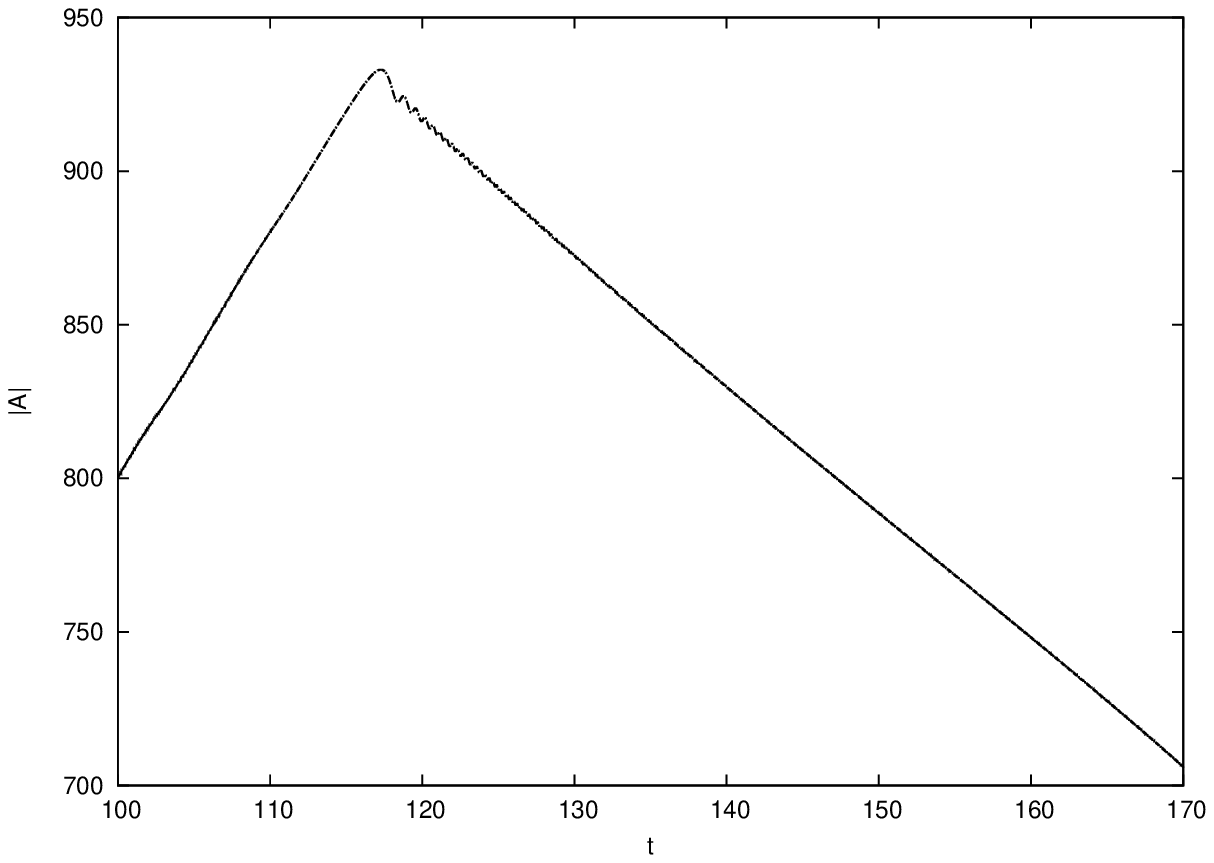}
\caption{График $|A(t)|$ при наличии диссипации. Коэффициенты диссипации $\mu_1=\mu_2=0.005$, величина амплитуды возмущения $f=18$.}
\label{fig2}
\end{figure}

\begin{figure}
\includegraphics[width=14cm,height=7cm]{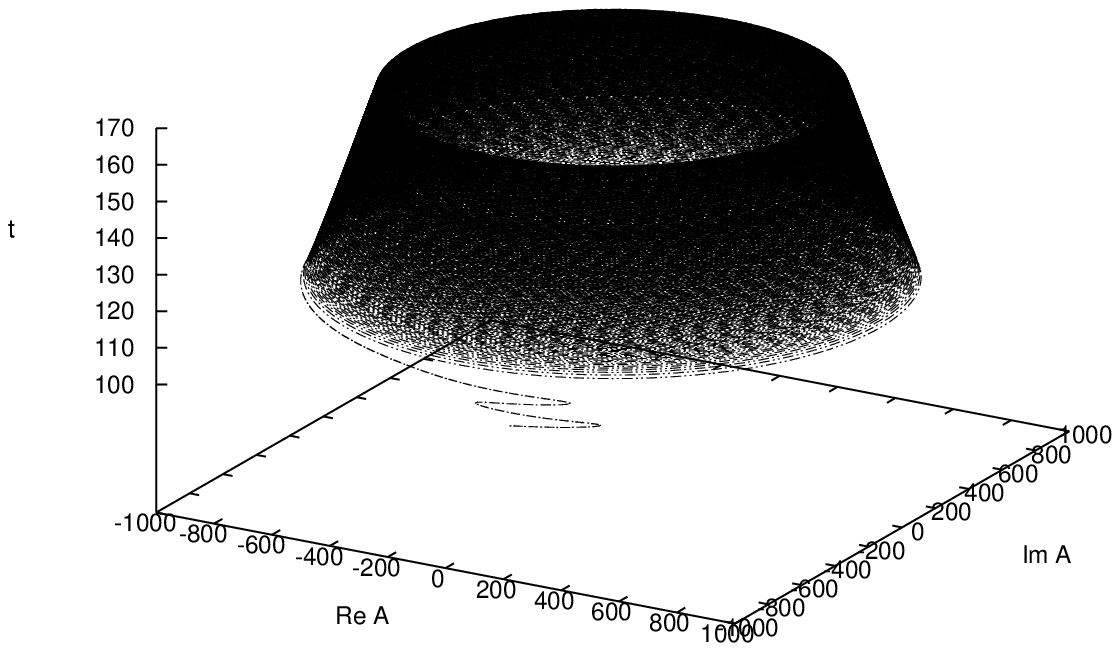}
\caption{Зависимость вещественной и мнимой части компоненты $A$ системы (\ref{mainSys}) от $t$. Коэффициенты диссипации $\mu_1=\mu_2=0.005$, величина амплитуды возмущения $f=18$.}
\label{fig3}
\end{figure}

На Рис. \ref{fig2}, \ref{fig3} приведены результаты численного счета решений системы (\ref{mainSys}) при наличии диссипации. Значения параметров -- $f=18, \mu_1=\mu_2=0.005$, величина шага сетки $10^{-4}$. 

На этих  рисунках отчетливо видны три типа движений. На начальном этапе происходит рост решения. На втором этапе --- происходит остановка роста из-за наличия диссипации в системе и наконец на третьем этапе решение начинает убывать.

Нашей задачей является получение асимптотических формул, описывающих данный эффект.

\section{Результат}

В работе предполагается, что коэффициенты диссипации $\mu_1, \mu_2$ в системе (\ref{mainSys}) имеют одинаковый порядок малости $\mu$ и мы используем обозначения $\mu_1=\delta_1\mu, \mu_2=\delta_2\mu; \delta_1, \delta_2 \sim O(1).$

Решение системы исходной системы (\ref{mainSys}) на начальном этапе авторезонансного роста имеет различное представление в разных областях. В области 
$$
\sqrt{f^2-16(3+2\delta_1\mu t+\delta_2\mu t)^2} \gg \mu 
$$
оно  имеет вид
\begin{eqnarray*}
A(t,\mu) &=&\frac{1}{\mu} \Big[8\theta + \mu^2 \alpha_2(\theta,\mu) \Big]\exp\left\{i \Psi_a(\theta,\mu) \right\},\nonumber \\
A(t,\mu) &=& \frac{1}{\mu} \Big[-4\theta + \mu^2 \beta_2(\theta,\mu) \Big]\exp\left\{2i \Psi_b(\theta,\mu)\right\}, \label{asypmAlgSol}
\end{eqnarray*}
где
\begin{eqnarray*}
\Psi_a &=& \Psi_0(\theta)+\mu^2\Psi_{a2}(\theta), \\
\Psi_b &=& \Psi_0(\theta)+\mu^2\Psi_{b2}(\theta),
\end{eqnarray*}
и независимая переменная $\theta=\mu t$.

В области 
$$
-\mu^{-1} \ll \frac{\sqrt{f^2-16(3+2\delta_1\mu t + \delta_2\mu t)^2}}{\mu} \ll \mu^{-1/2}
$$
решение имеет вид
\begin{eqnarray*}
a(\sigma,\mu) &=& \Big[a_0(\sigma)+\mu a_1(\sigma)+\mu^2 a_2(\sigma)\Big]\exp\Big\{i(\psi_0(\sigma)+\mu\psi_1(\sigma)+\mu^2\psi_2(\sigma))\Big\},\nonumber \\
b(\sigma,\mu) &=& \Big[b_0(\sigma)+\mu b_1(\sigma)+\mu^2 b_2(\sigma)\Big]\exp\Big\{2i(\psi_0(\sigma)+\mu\psi_1(\sigma)+\mu^2\psi_2(\sigma))\Big\}.\nonumber
\end{eqnarray*}
Здесь используется независимая переменная 
$$
\sigma=\frac{\sqrt{f^2-16(3+2\delta_1\theta+\delta_2\theta)^2}}{\mu}.
$$

Приведенные  коэффициенты асимптотических представлений определяются в разделах 5,6 соответственно.

\section{Алгебраические асимптотические решения}

В этом разделе мы будем исследовать  алгебраические решения системы (\ref{mainSys}) 
при значениях $t$, соответствующих начальному этапу авторезонансного роста решения.

Сделаем замену переменных 
$$
\theta=\mu t, \quad a(\theta,\mu)=\mu A, \quad b(\theta,\mu)=\mu B
$$

В новых переменных система (\ref{mainSys}) имеет вид

\begin{eqnarray}
\mu^2 a' &=& -i \big(2\theta a +\frac{1}{2}a^*b + \mu^2f \big) - \mu^2\delta_1a,\nonumber \\
\mu^2 b' &=& -i \big(4\theta b + \frac{1}{4}a^2 \big) - \mu^2\delta_2b, \label{mainSysAlg}
\end{eqnarray}
здесь мы предполагаем, что коэффициенты диссипации $\mu_1, \mu_2$ имеют один порядок $\mu$.

Решение (\ref{mainSysAlg})  будем искать в виде
\begin{eqnarray}
a(t,\mu) &=& \Big[8\theta + \mu^2 \alpha_2(\theta,\mu) + \mu^4 \alpha_4(\theta,\mu) \Big]\exp\left\{i \Psi_a(\theta,\mu) \right\},\nonumber \\
b(t,\mu) &=& \Big[-4\theta + \mu^2 \beta_2(\theta,\mu)+ \mu^4 \beta_4(\theta,\mu) \Big]\exp\left\{2i \Psi_b(\theta,\mu)\right\}, \label{asypmAlgSol}
\end{eqnarray}
где
\begin{eqnarray*}
\Psi_a &=& \Psi_0(\theta)+\mu^2\Psi_{a2}(\theta)+\mu^4\Psi_{a4}(\theta), \\
\Psi_b &=& \Psi_0(\theta)+\mu^2\Psi_{b2}(\theta)+\mu^4\Psi_{b4}(\theta).
\end{eqnarray*}

В представлении (\ref{asypmAlgSol}) в качестве главных членов асимптотики используются значения полученные при исследовании невозмущенной задачи из работы \cite{klg}.

Подставляя представление (\ref{asypmAlgSol}) в систему (\ref{mainSysAlg}) и собирая слагаемые при одинаковых степенях $\mu$ получим рекуррентную последовательность задач для определения коэффициентов $a_k, b_k, \psi_k$.

Соотношения при $\mu^0$ выполнены тождественно в силу выбора главных членов асимптотики. 

Для определения следующих поправок мы получаем систему линейных алгебраических уравнений с вырожденной матрицей. Соотношения при $\mu^2$ для определения $(\alpha_2,\beta_2,\Psi_{a2},\Psi_{b2})^T$ дают систему с матрицей 
$$
\left(
\begin{array}{cccc}
	1 & 1 & 0 &0 \\
	0 & 0 & -1 & 1 \\
	0& -\sin\Psi_0 & -8\theta\cos\Psi_0 & 8\theta\cos\Psi_0 \\
	0 & \cos\Psi_0 & -8\theta\sin\Psi_0 & 8\theta\sin\Psi_0
\end{array}
\right)
$$
Ее ранг равен трем и для разрешимости системы нужно требовать ортогональности правой части и решения союзной системы. 

Ортогональность решения союзной системы 
$$
\lambda_2\left(0,-\frac{8\theta}{\sin\Psi_0},\frac{\cos\Psi_0}{\sin\Psi_0},1\right)^T.
$$
и правой части 
\begin{eqnarray*}
\Big(2\Psi_0', \frac{1+\delta_2\theta}{8\theta^2},-[2\cos\Psi_0(1+\delta_1\theta)+2\theta\sin\Psi_0 \Psi_0']/\theta, \\
-[f+8\sin\Psi_0(1+\delta_1\theta)-8\theta\cos\Psi_0 \Psi_0']/4\theta\Big)^T
\end{eqnarray*}
 позволяет определить главный член $\Psi_0$
\bb
\sin\Psi_0 = - \frac{4(3+2\delta_1\theta+\delta_2\theta)}{f} \label{sinPsi0}.
\ee
Откуда получаем
$$
\Psi_0^{\prime} = - \frac{4(2\delta_1+\delta_2)}{\sqrt{f^2-16(3+2\delta_1\theta+\delta_2\theta)^2}}.
$$
Выражение (\ref{sinPsi0}) при $\delta_1=\delta_2=0$ совпадает с главным членом асимптотики невозмущенной задачи из работы \cite{klg}.

Соотношения при $\mu^4$ для определения $(\alpha_4,\beta_4,\Psi_{a4},\Psi_{b4})^T$ дают систему с матрицей 
$$
\left(
\begin{array}{cccc}
	1 & 1 & 0 &0 \\
	0 & 0 & -1 & 1 \\
	0& 0 & -1 & 1 \\
	0 & 1 & 0 & 0
\end{array}
\right)
$$
Ее ранг также равен  трем и для разрешимости системы нужно требовать ортогональности правой части и решения союзной системы. 

Решение союзной системы имеет вид
$$
\lambda_4 \left(0, 1, 1, 0 \right)^T
$$

Ее ортогональность к правой части $(f_1,f_2,f_3,f_4)$, где
\begin{eqnarray*}
f_1=\Big[-\alpha_2^2+128\theta^2\Psi_{a2}^2+32\Psi_{b2}+32\delta_2\theta\Psi_{b2}-128\theta^2\Psi_{b2}^2-8\beta_2\Psi_0'+ \\
32\theta\Psi_{b2}' \Big]/(16\theta),\\
f_2=\left[4\delta_2\beta_2-32\theta\alpha_2\Psi_{a2}-32\theta\beta_2\Psi_{b2}+2\beta_2'+64\theta\Psi_{b2}\Psi_0' \right]/(128\theta^2),\\
f_3=\Big[-2\delta_1\alpha_2+8\theta\alpha_2\Psi_{a2}+8\theta\beta_2\Psi_{a2}-8\theta\alpha_2\Psi_{b2}+16\theta\beta_2\Psi_{b2}+ \\
16\theta\Psi_{a2}\Psi_0' \Big]/(64\theta^2),\\
f_4= \Big[-\alpha_2\beta_2-16\Psi_{a2}-16\delta_1\theta\Psi_{a2}+64\theta^2\Psi_{a2}\Psi_{b2}-64\theta^2\Psi_{b2}^2 - \\
2\alpha_2\Psi_0'-16\theta\Psi_{a2}' \Big]/(8\theta).
\end{eqnarray*}
позволяет определить поправки $\alpha_2, \beta_2$. В результате получаем
\begin{eqnarray}
\alpha_2 &=& \frac{f^2-16(3+2\delta_1\theta+\delta_2\theta)^2-48(2\delta_1+\delta_2)}{4\sqrt{f^2-16(3+2\delta_1\theta+\delta_2\theta)^2}},\nonumber\\
\beta_2 &=& \frac{16(3+2\delta_1\theta+\delta_2\theta)^2+32(2\delta_1+\delta_2)-f^2}{4\sqrt{f^2-16(3+2\delta_1\theta+\delta_2\theta)^2}}.\label{sol2ext}
\end{eqnarray}

Полученное решение (\ref{sol2ext}) для  второй поправки амплитуды позволяет определить область пригодности представления (\ref{asypmAlgSol}):
\bb
\sqrt{f^2-16(3+2\delta_1\theta+\delta_2\theta)^2}\mu^{-1}\gg 1.
\ee

\begin{figure}
\includegraphics[width=7cm,height=13cm,angle=-90]{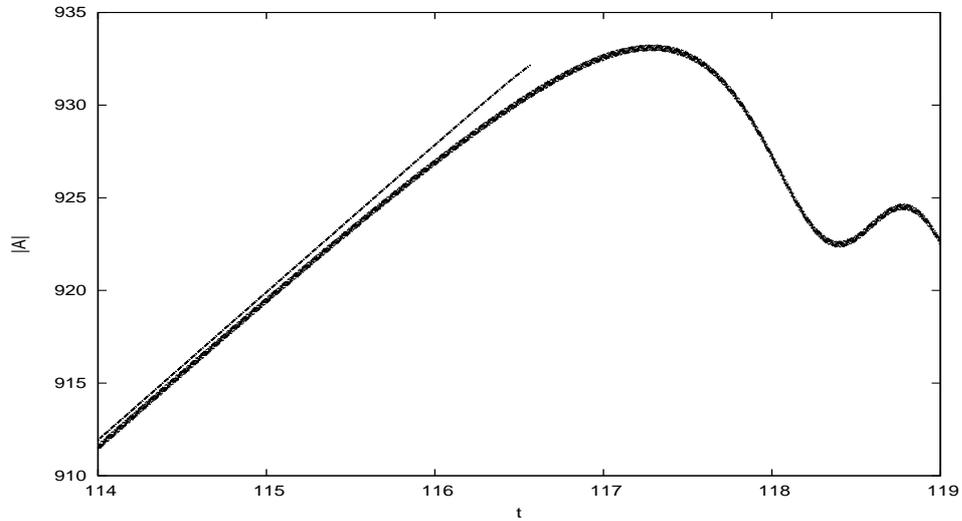}
\caption{Асимптотика (\ref{asypmAlgSol}) (тонкая линия) и численное решение (толстая линия). При численном счете использовались коэффициенты диссипации $\delta_1=\delta_2=1, \mu=0.005$, величина амплитуды возмущения $f=18$. Остановка роста тонкой  линии соответствует выходу из области пригодности для представления (\ref{asypmAlgSol})}
\label{fig4}
\end{figure}

Численный счет позволяет проверить построенное асимптотическое решение. На рисунке \ref{fig4} приведено результаты  для численного и асимптотического решения (\ref{asypmAlgSol}). Тонкая линия соответствует асимптотическому решению, толстая --- численному. Численное решение получено методом Рунге-Кутты четвертого порядка. Величина шага сетки составляла $10^{-4}$. Значения параметров $\delta_1=\delta_2=1, \mu=0.005$.
Величина амплитуды возмущения $f=18$, что соответствует наличию авторезонанса в системе, см. \cite{klg}.

\begin{figure}
\includegraphics[width=7cm,height=13cm,angle=-90]{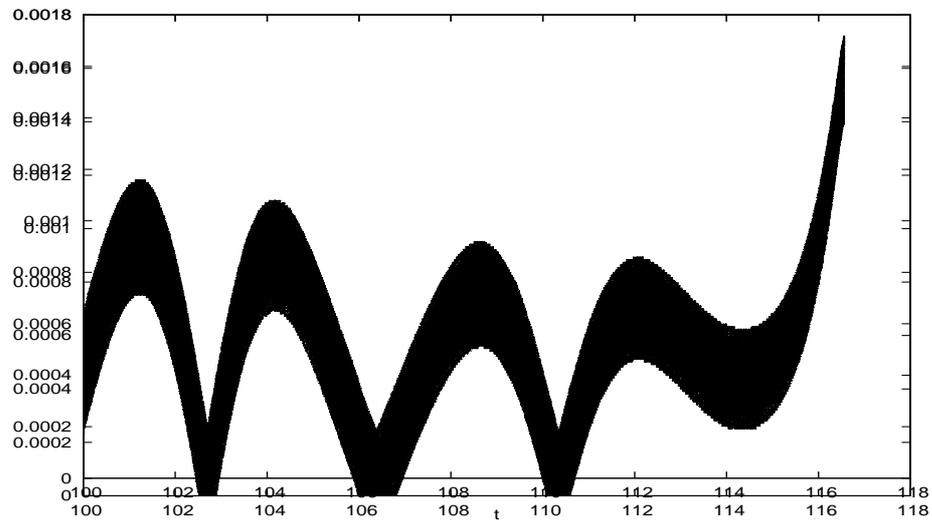}
\caption{Относительная разность $|a_{num}(t)-a(t)|/|a_{num}(t)|$ для компоненты $a$ представления (\ref{asypmAlgSol}). Коэффициенты диссипации $\delta_1=\delta_2=1, \mu=0.005$, величина амплитуды возмущения $f=18$.}
\label{fig5}
\end{figure}

На рисунке \ref{fig5} мы приводим относительную разность \linebreak 
$|a_{num}(t)-a(t)|/|a_{num}(t)|$ для представления (\ref{asypmAlgSol}) компоненты $a$ решения системы (\ref{mainSysAlg}). Здесь $a$ --- асимптотическое решение, $a_{num}$ --- численное решение.

\section{Окрестность обрыва авторезонансного роста}

Область пригодности асимптотического представления (\ref{asypmAlgSol}) полученная в предыдущем  параграфе определяет новую растянутую внутреннюю переменную. В исходной системе уравнений перейдем к переменной 
$$
\sigma=\frac{S}{\mu},\quad \hbox{где}\quad S=\sqrt{f^2-16(3+2\delta_1\theta+\delta_2\theta)^2}.
$$

В новой переменной $\sigma$ исследуемая система уравнений (\ref{mainSysAlg}) примет вид
\begin{eqnarray}
\frac{8\sqrt{f^2-\mu^2\sigma^2}(2\delta_1+\delta_2)}{\sigma}a'_\sigma &=& i\left[\frac{\sqrt{f^2-\mu^2\sigma^2}-12}{2(2\delta_1+\delta_2)}a+\frac{1}{2}a^*b+\mu^2f \right]-\mu^2\delta_1 a, \nonumber
\\
\frac{8\sqrt{f^2-\mu^2\sigma^2}(2\delta_1+\delta_2)}{\sigma}b'_\sigma &=& i\left[\frac{\sqrt{f^2-\mu^2\sigma^2}-12}{(2\delta_1+\delta_2)}b+\frac{1}{4}a^2 \right]-\mu^2\delta_2 b.\label{neibSys}
\end{eqnarray}

Решение будем искать в виде

\begin{eqnarray}
a(\sigma,\mu) &=& \Big[a_0(\sigma)+\mu a_1(\sigma)+\mu^2 a_2(\sigma)\Big]\exp\Big\{i(\psi_0(\sigma)+\mu\psi_1(\sigma)+\mu^2\psi_2(\sigma))\Big\},\nonumber \\
\label{intExp1}\\
b(\sigma,\mu) &=& \Big[b_0(\sigma)+\mu b_1(\sigma)+\mu^2 b_2(\sigma)\Big]\exp\Big\{2i(\psi_0(\sigma)+\mu\psi_1(\sigma)+\mu^2\psi_2(\sigma))\Big\}.\nonumber
\end{eqnarray}

Подставляя представление (\ref{intExp1}) в систему (\ref{neibSys}) и собирая слагаемые при одинаковых степенях малого параметра $\mu$ мы получим рекуррентную последовательность уравнений для определения коэффициентов представления (\ref{intExp1}).

Система (\ref{neibSys}) в главном имеет вид

\begin{eqnarray}
\frac{8f(2\delta_1+\delta_2)}{\sigma}\Big[a'_0 + i\psi'_0 a_0\Big] &=& i\left[\frac{f-12}{2(2\delta_1+\delta_2)}a_0+\frac{1}{2}a_0b_0 \right], \nonumber\\
\frac{8f(2\delta_1+\delta_2)}{\sigma}\Big[b'_0+2i\psi'_0 b_0 \Big]&=& i\left[\frac{f-12}{(2\delta_1+\delta_2)}b_0+\frac{1}{4}a_0^2 \right].\label{intMainSys}
\end{eqnarray}

Анализируя систему получаем, что главные члены амплитуд $a_0, b_0$ являются постоянными и связаны между собой соотношением $a_0^2=4b_0^2$. На этом же этапе выводится уравнение для главного члена $\psi_0$ фазы 
$$
\psi'_0(\sigma)=\frac{f-12+b_0(2\delta_1+\delta_2)}{16f(2\delta_1+\delta_2)^2}\sigma.
$$

Интегрируя его получаем

$$
\psi_0(\sigma)=\frac{f-12+b_0(2\delta_1+\delta_2)}{32f(2\delta_1+\delta_2)^2}\sigma^2+\psi_{0,0}.
$$

В результате главные члены представления (\ref{intExp1}) имеют вид
\begin{eqnarray}
a \sim 2b_0\exp\left\{i\left[\psi_{0,0}+\frac{f-12+\rho_0(2\delta_1+\delta_2)}{32f(2\delta_1+\delta_2)^2}\sigma^2\right]\right\},\nonumber\\ b \sim b_0\exp\left\{2i\left[\psi_{0,0}+\frac{f-12+\rho_0(2\delta_1+\delta_2)}{32f(2\delta_1+\delta_2)^2}\sigma^2\right]\right\},\label{sol}
\end{eqnarray}

где постоянные $b_0, \psi_{0,0}$ определяются согласованием с представлением (\ref{asypmAlgSol})
\bb
b_0 = \frac{f-12}{2\delta_1+\delta_2},\qquad \psi_{0,0} = -\frac{\pi}{2}.
\ee
На следующем шаге из соотношений при первой степени малого параметра $\mu$ получаем систему
\begin{eqnarray}
\frac{8f(2\delta_1+\delta_2)}{\sigma}\Big[a'_1+i\psi'_0a_1+i\psi'_1a_0-\psi'_0\psi_1a_0  \Big]&=& \nonumber\\ i\Big[\frac{f-12}{2(2\delta_1+\delta_2)}(a_1+ia_0\psi_1)&+&\frac{1}{2}(a_1b_0+a_0b_1+i\psi_1a_0b_0)\Big],\nonumber\\ 
\frac{8f(2\delta_1+\delta_2)}{\sigma}\Big[b'_1+2i\psi'_0b_1+2i\psi'_1b_0-4\psi'_0\psi_1b_0  \Big]&=& \nonumber\\ i\Big[\frac{f-12}{(2\delta_1+\delta_2)}(b_1+2ib_0\psi_1)&+&\frac{1}{2}(a_0a_1+i\psi_1a^2_0)\Big]
\end{eqnarray}

Группируя вещественные и мнимые слагаемые слагаемые определяем, что коэффициенты $a_1, b_1$ являются постоянными и связаны соотношением $a_1=2b_1$. Поправка $\psi_1$ в фазе удовлетворяет уравнению
\bb
\psi'_1(\sigma)=\frac{b_1}{16f(2\delta_1+\delta_2)}\sigma.
\ee

Соотношения при $\mu^2$, после подстановки уже найденных предыдущих коэффициентов представления (\ref{intExp1}) имеют вид
\begin{eqnarray}
16if(2\delta_1+\delta_2)a'_2-32(2\delta_1+\delta_2)fb_0\psi'_2 = 2f\sigma\Big(i\sin\psi_0-\cos\psi_0\Big)+4i\delta_1b_0\sigma -  \nonumber\\
2b_0b_2\sigma +\frac{b_0\Big(12-b_0(2\delta_1+\delta_2)\Big)\sigma^3}{f^2(2\delta_1+\delta_2)}, \nonumber\\
4f(2\delta_1+\delta_2)\Big(\cos\psi_0+i\sin\psi_0\Big)\Big[8f(2\delta_1+\delta_2)b'_2+16ifb_0(2\delta_1+\delta_2)\psi'_2\Big] = \nonumber\\
-4f(2\delta_1+\delta_2)\Big(\cos\psi_0+i\sin\psi_0\Big)\Big[\delta_2b_0\sigma-ib_0a_2\sigma+ib_0b_2\sigma \Big]+ \nonumber\\ \frac{2b_0(-12+b_0(2\delta_1+\delta_2))(i\cos\psi_0-\sin\psi_0)\sigma^3}{f}.\label{eqMu2}
\end{eqnarray}

Выделение слагаемых для вещественной и мнимой части из системы (\ref{eqMu2}) дает уравнения для определения вторых поправок
\begin{eqnarray}
a'_2(\sigma)=\frac{f\sin\psi_0+2\delta_1b_0}{8f(2\delta_1+\delta_2)}\sigma, \hskip7cm  \nonumber\\
b'_2(\sigma)=-\frac{b_0\sigma}{32f^3(2\delta_1+\delta_2)^2\sin\psi_0} 
\Big\{\Big[2f^2\delta_2(2\delta_1+\delta_2)-2a_2f^2(2\delta_1+\delta_2) + \nonumber\\ 2b_2f^2(2\delta_1+\delta_2)-12\sigma^2+b_0(2\delta_1+\delta_2)\sigma^2\Big] \cos\psi_0 
+ \Big[2f^2\delta_2(2\delta_1+\delta_2) + \nonumber\\
2a_2f^2(2\delta_1+\delta_2) -  2b_2f^2(2\delta_1+\delta_2) -12\sigma^2+b_0(2\delta_1+\delta_2)\sigma^2\Big]\sin\psi_0 \Big\},\nonumber \\
\psi'_2(\sigma)=-\frac{\sigma}{64f^3(2\delta_1+\delta_2)^2\cos\psi_0} 
\Big\{\Big[-2f^2\delta_2(2\delta_1+\delta_2)-2a_2f^2(2\delta_1+\delta_2) + \nonumber\\ 2b_2f^2(2\delta_1+\delta_2)-12\sigma^2 + b_0(2\delta_1+\delta_2)\sigma^2\Big] \cos\psi_0 
+ \Big[2f^2\delta_2(2\delta_1+\delta_2) - \nonumber\\
2a_2f^2(2\delta_1+\delta_2) +  2b_2f^2(2\delta_1+\delta_2) + 12\sigma^2 - b_0(2\delta_1+\delta_2)\sigma^2\Big]\sin\psi_0 \Big\}.\label{eq2corr}
\end{eqnarray}

В результате поведение коэффициентов разложения амплитуды приводит к определению области пригодности представления (\ref{intExp1})
$$
\sigma \ll \mu^{-1/2}.
$$

\end{document}